# ON MEASURING STANDARDS

# IN

# WEYL'S GEOMETRY

Mark Israelit

Department of Physics and Mathematics, University of Haifa-Oranim,
Tivon 36006 Israel.  E-mail: <israelit@macam.ac.il>

*In Weyl's geometry the nonintegrability problem and difficulties in defining measuring standards are reconsidered. Approaches removing the nonintegrability of length in the interior of atoms are given, so that atoms can serve as measuring standards. The Weyl space becomes a well founded framework for classical theories of electromagnetism and gravitation.*

# 1. INTRODUCTION

Recently [1] it was proposed to enrich the 5-dimensional bulk of Wesson's induced matter theory [2] by the Weyl-Dirac geometric framework [3]. On the 4-dimensional brane this generalization results in a Weyl-type theory of gravitation and electromagnetism with mass and electric current induced by the geometry of the bulk. However, in Weyl's geometry [4] one is faced by the nonintegrability of length, that causes difficulties in determining measuring standards. This obstacle was emphasized by Einstein [5] some years after Weyl had proposed his theory. In the present note we reconsider the nonintegrability problem in Weyl's geometry and describe some procedures of canceling the nonintegrability in the interior of measuring standards. As a result the Weyl-Dirac generalization [1] of Wesson's theory [2] becomes an acceptable framework for describing phenomena of classical relativistic physics.

# 2. THE MEASURING PROBLEM IN WEYL'S GEOMETRY

In order to enlighten the problem we begin with a concise summary of details relevant to the following discussion. We deal with Weyl's geometry as modified by Dirac [6], also known as the Weyl–Dirac (W-D) framework. A detailed description of the W-D theory may be found in the works of Weyl [4], Dirac [6], Rosen [12], as well in [3].

Weyl [4] issued from the idea that by parallel displacement both, the component of a vector and its length, change. Thus, under an infinitesimal parallel displacement $dx^\nu$ of a given vector, its component $V^\mu$ changes according to

$$dV^\mu = -V^\sigma \Gamma^\mu_{\sigma\nu} dx^\nu \ , \tag{1}$$

while the change in the length, $V = (g_{\lambda\sigma} V^\lambda V^\sigma)^{\frac{1}{2}}$, of the same vector is

$$dV = V w_\nu dx^\nu \ . \tag{2}$$

In Eqs. (1), (2) $\Gamma^\mu_{\sigma\nu}$ is the affine connection, and $w_\nu$ is Weyl's length connection vector. Assuming that in every point of the space-time exist the metric tensor, $g_{\mu\nu} = g_{\nu\mu}$, and $w_\nu$, Weyl derived from (1) and (2) the following relation for the affine connection:

$$\Gamma^\lambda_{\mu\nu} = \left\{{}^{\lambda}_{\mu\nu}\right\} + g_{\mu\nu} w^\lambda - \delta^\lambda_\mu w_\nu - \delta^\lambda_\nu w_\mu \ . \tag{3}$$

Here $\left\{{}^{\lambda}_{\mu\nu}\right\}$ stands for the Christoffel symbol. Now, if the vector $V^\mu$ is transported by parallel displacement round an infinitesimal parallelogram with sides $dx^\mu$ and $\delta x^\nu$, one has from (1) the change of the component

$$\Delta V^\lambda = V^\sigma K^\lambda_{\sigma\mu\nu} dx^\mu \delta x^\nu . \tag{4}$$

In equation (4) $K^\lambda_{\sigma\mu\nu}$ stands for the curvature tensor of the W-D geometry, which is given by an expression like the Riemannian curvature tensor, but with the connection $\Gamma^\lambda_{\mu\nu}$ instead of $\left\{{}^{\lambda}_{\mu\nu}\right\}$. From (2) one obtains for the change in the length of the vector that was transported round the parallelogram

$$\Delta V = V W_{\mu\nu} dx^\mu \delta x^\nu , \tag{5}$$

where $W_{\mu\nu} \equiv w_{\mu,\nu} - w_{\nu,\mu}$ is the Weyl length curvature tensor. If a vector $V$ has been transported round a closed loop, and arrived at the starting point, its new length according to (5) is given by

$$V_{new} = V_{initial} + \int_S V W_{\mu\nu} dS^{\mu\nu}, \tag{6}$$

with $S$ being the area of the loop and $dS^{\mu\nu}$ an element of this area. One concludes that the length is nonintegrable, unless $w_\nu$ is a gradient vector, so that $W_{\mu\nu} = 0$.

From (6) one sees that the length of a vector is not simply determined, so that he can choose an arbitrary standard of length, or a gauge, at each point. Thus, one can introduce a local gauge transformation, the Weyl gauge transformation (WGT), as follows. If under WGT the component $V^\mu$ of a vector remains unchanged, the length of this vector changes according to

$$V \Rightarrow \tilde{V} = e^\lambda V, \tag{7}$$

where $\lambda(x^\nu)$ is an arbitrary function. Under the WGT (7), the metric tensor changes as

$$g_{\mu\nu} \Rightarrow \tilde{g}_{\mu\nu} = e^{2\lambda} g_{\mu\nu}, \quad \text{and} \quad g^{\mu\nu} \Rightarrow \tilde{g}^{\mu\nu} = e^{-2\lambda} g^{\mu\nu}, \tag{8}$$

and the Weylian connection vector changes according to

$$w_\nu \Rightarrow \tilde{w}_\nu = w_\nu + \lambda_{,\nu}. \tag{9}$$

Modifying the Weyl theory, Dirac [6] introduced into the framework a function $\beta(x^\mu)$ - the Dirac gauge function that under WGT transforms as follows

$$\beta \Rightarrow \tilde{\beta} = e^{-\lambda} \beta. \tag{10}$$

It must be pointed that there is a one-to-one correspondence between $\lambda(x^\mu)$ and the Dirac gauge function $\beta(x^\mu)$. Indeed, one can always fix $\beta(\lambda = 0) = 1$, so that $\beta = e^{-\lambda}$. In a Weyl-Dirac space one can carry out both, coordinate transformations (CT), - as in the Riemann geometry, - and WGT.

Following Dirac [6] we introduce the concept of a gauge covariant quantity and its Weyl power. If under WGT a quantity $\Psi^{\alpha..\beta}_{\mu..\nu}$ labeled by coordinate indices is transformed according to the law

$$\Psi^{\alpha..\beta}_{\mu..\nu} \Rightarrow \tilde{\Psi}^{\alpha..\beta}_{\mu..\nu} = e^{n\lambda}\Psi^{\alpha..\beta}_{\mu..\nu}, \qquad (11)$$

it is called a co-quantity of power $n$. If $n = 0$, then $\Psi^{\alpha..\beta}_{\mu..\nu}$ is an in-quantity. Following Canuto *et al.* [7] we will denote the Weyl power by $\Pi$, so that $\Pi(g_{\mu\nu}) = 2$, $\Pi(g^{\mu\nu}) = -2$, $\Pi(\beta) = -1$ and for the vector considered in (7) $\Pi(V^\mu) = 0$, $\Pi(V) = 1$. One can also verify that the connection (3) is gauge invariant i.e. $\Pi(\Gamma^\lambda_{\mu\nu}) = 0$.

Let us turn to the measuring problem. In Weyl's geometry one has in every point an arbitrary gauge and hence arbitrary lengths. However, even in this case the ratio of lengths is well determined and one can still measure the dimensions of physical objects. The real difficulty in Weyl's geometry is the nonintegrability of length rather than the arbitrary gauge. Having in mind a typical example of measuring lengths by means of light waves emitted by an atom, we consider the following *Gedankenversuch*. Assume there are two identical atoms at a point *P* of space-time, while at a second point *Q* we have a physical object *<Obj>*. In order to measure the properties of *<Obj>* one carries the atoms from *P* to *Q*. Now, if the atoms are brought to *Q* by different paths they will no longer be identical; they will, in general, have different properties (cf. (6)). Thus, one has no longer a standard.

Dirac [6] assumed that in practice one makes use of two different intervals: $ds_A$ and $ds_E$. The interval $ds_A$ is referred to atomic units; it does not depend on an arbitrary metric gauge and is not affected by WGT. The Einstein interval $ds_E$ is associated with the field equations and the Weyl geometry, so that it is nonintegrable under a parallel displacement, and in order to get a definite value of $ds_E$ a certain gauge must be chosen.

In the following sections we consider some possible geometric structures for the interior of atoms serving as Dirac's measuring standards.

## 3. THE BUBBLE MODEL

An interesting solution of the problem was given by Wood and Papini [8]. In their model the atom, serving for a measuring standard, appears as a bubble. Outside one has the Weyl space with nonintegrability of length and WGT invariance; the Weyl connection vector $w_\nu(x^\mu)$ and the Dirac gauge function $\beta(x^\mu)$ determine this space together with the metric tensor $g_{\mu\nu}$. On the boundary surface and in the interior of the particle Wood and Papini took

$$w_\mu = 0, \tag{12}$$

and

$$\beta = \beta_0 = const. \tag{13}$$

The static spherically symmetric entity is filled with "Dirac matter" satisfying the equation of state

$$\rho = -P, \tag{14}$$

where $\rho$ stands for the matter density, and $P$ denotes the pressure. The interior of the bubble is homogeneous, and the matter density is given by

$$\rho = \frac{1}{2}\Lambda\beta_0{}^2, \tag{15}$$

with $\Lambda$ (which is regarded as an arbitrary constant) stemming from the cosmological term in Dirac's action integral (cf. [6], [3]). The procedure of Papini and Wood [8] is based on the concept of an infinitesimally thin shell of matter considered in a generalized Gauss-Mainardi-Codazzi formalism [9, 10]. In the exterior space one has the Weyl geometry with the freedom of both CT and WGT, and the electromagnetic field can be given a geometric interpretation. At the same time a standard of length can be introduced into the theory by breaking the Weylian gauge-invariance in the interior of the bubble.

## 4. NONMETRICITY OR TORSION

It is possible in Weyl's geometry to replace the nonmetricity by torsion [11]. By this procedure one can obtain an integrable space-time in the interior of the atom, so that its essential properties remain unchanged under displacements.

For a moment let us consider a more general space-time, which is characterized by the metric tensor $g_{\mu\nu} = g_{\nu\mu}$ and by a connection $\hat{\Gamma}^\lambda_{\mu\nu}$, that may differ from the Weylian connection (3). We introduce the nonmetricity tensor

$$\hat{Q}_{\lambda\mu\nu} = -\hat{\nabla}_\lambda g_{\mu\nu} \equiv -\frac{\partial g_{\mu\nu}}{\partial x^\lambda} + g_{\sigma\nu}\hat{\Gamma}^\sigma_{\mu\lambda} + g_{\mu\sigma}\hat{\Gamma}^\sigma_{\nu\lambda}, \tag{16}$$

and the torsion tensor

$$\hat{\Gamma}^\lambda_{[\mu\nu]} = \frac{1}{2}\left(\hat{\Gamma}^\lambda_{\mu\nu} - \hat{\Gamma}^\lambda_{\nu\mu}\right). \tag{17}$$

Further, we write down the contorsion tensor given by

$$\hat{C}^{\lambda}_{\mu\nu} \equiv \hat{\Gamma}^{\lambda}_{[\mu\nu]} + g^{\lambda\sigma}g_{\mu\rho}\hat{\Gamma}^{\rho}_{[\nu\sigma]} + g^{\lambda\sigma}g_{\nu\rho}\hat{\Gamma}^{\rho}_{[\mu\sigma]}. \tag{18}$$

In terms of nonmetricity and contorsion the connection $\hat{\Gamma}^{\lambda}_{\mu\nu}$ may be expressed as follows (cf. [13, 14])

$$\hat{\Gamma}^{\lambda}_{\mu\nu} = \left\{{}^{\lambda}_{\mu\nu}\right\} + \hat{C}^{\lambda}_{\mu\nu} + \frac{1}{2}g^{\lambda\sigma}\left(\hat{Q}_{\nu\mu\sigma} + \hat{Q}_{\mu\nu\sigma} - \hat{Q}_{\sigma\mu\nu}\right). \tag{19}$$

One can easily verify that nonintegrability is caused by nonmetricity. Going back to the proper Weyl case described in brief in section 2 one obtains from (3)

$$\Gamma^{\lambda}_{[\mu\nu]} = 0; \quad C^{\lambda}_{\mu\nu} = 0; \quad Q_{\lambda\mu\nu} = -2g_{\mu\nu}w_{\lambda}, \tag{20}$$

i.e. a torsionless geometry with nonmetricity.

Below we will show that the nonintegrable Weyl framework may be replaced by an integrable geometry with torsion. In the procedure we will make use of an approach proposed by Nathan Rosen [12].

Let us issue from the connection (3) and calculate the change in length of a vector under an infinitesimal displacement. For the contravariant component we have equation (1), whereas the change of the covariant component may be accounted as

$$dV_{\mu} = d\left(g_{\mu\sigma}V^{\sigma}\right) = g_{\mu\sigma}dV^{\sigma} + V^{\sigma}g_{\mu\sigma,\nu}dx^{\nu}. \tag{21}$$

Substituting (1) and (3) into (21) yields

$$dV_{\mu} = V_{\lambda}\left(\left\{{}^{\lambda}_{\mu\nu}\right\} + g_{\mu\nu}w^{\lambda} + \delta^{\lambda}_{\mu}w_{\nu} - \delta^{\lambda}_{\nu}w_{\mu}\right)dx^{\nu} = V_{\lambda}\underset{2}{\Gamma}^{\lambda}_{\mu\nu}dx^{\nu}. \tag{22}$$

One can calculate the change of the length according to the rule $dV^2 \equiv d(V_{\mu}V^{\mu}) = V_{\mu}dV^{\mu} + V^{\mu}dV_{\mu}$, and make use of (1) and (22). Then one obtains the result given in (2). For the change in the length of the vector after being transported round a parallelogram one obtains of course (5).

Now, besides the connection (3) one has from (22) a new connection

$$\underset{2}{\Gamma}^{\lambda}_{\mu\nu} = \left\{{}^{\lambda}_{\mu\nu}\right\} + g_{\mu\nu}w^{\lambda} + \delta^{\lambda}_{\mu}w_{\nu} - \delta^{\lambda}_{\nu}w_{\mu}. \tag{23}$$

This has a nonzero torsion $\Gamma^{\lambda}_{2\,[\mu\nu]} = \delta^{\lambda}_{\mu} w_{\nu} - \delta^{\lambda}_{\nu} w_{\mu}$ and a nonmetricity $Q_{2\,\lambda\mu\nu} = 2 g_{\mu\nu} w_{\lambda}$, which is opposite to that given in (20).

There is an alternative way of considering the change in length of a vector. Instead of issuing from (1), one can start with the covariant component and make use of connection (3). Then the law of parallel displacement is

$$d_2 V_{\mu} = V_{\lambda} \Gamma^{\lambda}_{\mu\nu} dx^{\nu} . \tag{24}$$

For the change of the contravariant component, $V^{\mu} = g^{\mu\sigma} V_{\sigma}$ one obtains from (24)

$$d_2 V^{\mu} = -V^{\sigma} \Gamma^{\mu}_{2\,\sigma\nu} dx^{\nu} . \tag{25}$$

With (24) and (25) one obtains the change in length of the displaced vector

$$d_2 V = -V w_{\nu} dx^{\nu} . \tag{26}$$

Finally, for a vector transported round a parallelogram the change in length is

$$\Delta_2 V = -V W_{\mu\nu} dx^{\mu} \delta x^{\nu} . \tag{27}$$

To sum up, there are two alternatives: **1)** We can start with the law of parallel displacement (1) and with the connection given by (3). In this case a second connection (cf. (23)) appears. The change in the length of a vector is then given by (2) and (5). **2)** We can start with the law (25) and with connection (23). Then the original Weyl connection (3) is evoked. The length changes in this case according to equations (26) and (27). It is interesting that the changes in length, obtained by the two procedures, are opposite in sign. Thus, we can turn from (2) and (5) to the results given in (26) and (27) by changing the sign of the Weylian vector $w_{\nu}$.

The existence of two kinds of parallel displacement (1) and (25) justifies defining the following third kind of displacement:

$$\underset{3}{d}V^{\mu} = \frac{1}{2}\left(dV^{\mu} + \underset{2}{d}V^{\mu}\right) = -V^{\sigma}\underset{3}{\Gamma^{\mu}}_{\sigma\nu}dx^{\nu}, \tag{28}$$

with a new connection given by

$$\underset{3}{\Gamma^{\lambda}}_{\mu\nu} = \frac{1}{2}\left(\Gamma^{\lambda}_{\mu\nu} + \underset{2}{\Gamma^{\lambda}}_{\mu\nu}\right) = \left\{{}^{\lambda}_{\mu\nu}\right\} + g_{\mu\nu}w^{\lambda} - \delta^{\lambda}_{\nu}w_{\mu}. \tag{29}$$

One can readily prove that the nonmetricity vanishes

$$\underset{3}{Q}_{\lambda\mu\nu} = -\underset{3}{\nabla}_{\lambda}g_{\mu\nu} \equiv -\frac{\partial g_{\mu\nu}}{\partial x^{\lambda}} + g_{\sigma\nu}\underset{3}{\Gamma^{\sigma}}_{\mu\lambda} + g_{\mu\sigma}\underset{3}{\Gamma^{\sigma}}_{\nu\lambda} = 0, \tag{30}$$

and that the new connection has a torsion

$$\underset{3}{\Gamma^{\lambda}}_{[\mu\nu]} = \frac{1}{2}\left(\delta^{\lambda}_{\mu}w_{\nu} - \delta^{\lambda}_{\nu}w_{\mu}\right). \tag{31}$$

Adopting $\underset{3}{\Gamma}$ we obtain for the covariant components of the vector

$$\underset{3}{d}V_{\mu} = V_{\sigma}\underset{3}{\Gamma^{\sigma}}_{\mu\nu}dx^{\nu}, \tag{32}$$

so that the length of a vector remains unchanged under parallel displacement.

$$\underset{3}{d}V = 0 \qquad \text{and} \qquad \underset{3}{\Delta}V = 0. \tag{33}$$

In the W-D framework one describes gravitation by the metric tensor $g_{\mu\nu}$, while electromagnetism is introduced by means of the Weylian connection vector $w_{\mu}$, the latter being interpreted as the vector potential. If the geometry is given by $g_{\mu\nu}$ and by $\Gamma^{\lambda}_{\mu\nu}$ (cf. (3)), Maxwell's field tensor is identified with the Weyl length curvature tensor $W_{\mu\nu} = w_{\mu,\nu} - w_{\nu,\mu}$ (cf. (5)). Alternatively, if the geometry is described by $g_{\mu\nu}$ and by $\underset{3}{\Gamma^{\lambda}}_{\mu\nu}$ (cf. (29)), the space is integrable, and the Maxwell field tensor is given by the divergence of the torsion tensor

$$w_{\mu;\nu} - w_{\nu;\mu} = -2\underset{3}{\Gamma^{\lambda}}_{[\mu\nu];\lambda}. \tag{34}$$

Let us adopt the standpoint that the geometry in the interior of atoms differs from that describing the exterior space.

Then, the exterior may be described by Weyl's geometry with $g_{\mu\nu}$ and $\Gamma^{\lambda}_{\mu\nu}$ (cf. (3)), that possesses WGT, the latter giving a geometric interpretation for gauge transformations of the vector potential. This framework is a suitable basis for a geometrically based theory of gravitation and electromagnetism. The interior of the atom can be characterized by $g_{\mu\nu}$ and the connection $\Gamma^{\lambda}_{\mu\nu 3}$ (cf. (29)). Here torsion formed from $w_{\lambda}$ appears, but nonmetricity is cancelled. Thus, in the interior the length is integrable and both atoms in our *Gedankenversuch* (cf. section 2.) remain identical after the displacement.

It is interesting that the infinitesimal properties of the interior are affected by a contorsion tensor. This fact agrees with the well known conception of linking intrinsic microscopic properties of matter with contorsion (cf. [15]). Further, the interior and the exterior geometries are constructed from the same basic quantities $g_{\mu\nu}$ and $w_{\mu}$, so both can be treated as different representation of the same physical reality.

## 5. GAUGE-COVARIANT DISPLACEMENT AND STANDARD VECTORS

An elegant way of removing the nonintegrability obstacle was proposed by Nathan Rosen [12]. His procedure is based on gauge-covariant derivatives. Consider a coordinate-scalar $S(x)$ having the Weyl power $n$, so that after a WGT one has $\tilde{S} = e^{n\lambda} S$. The partial derivative of the transformed scalar is

$$\tilde{S}_{,\nu} = e^{n\lambda}\left(S_{,\nu} + nS\lambda_{,\nu}\right).  \tag{35}$$

This expression is obviously not covariant with respect to WGT, unless $n = 0$. Let us look for a differential operator that does not break the gauge invariance. Following Rosen [12] we can define a *gauge covariant* derivative of $S$

$$S_{|\nu} \equiv S_{,\nu} - nSw_{\nu}.  \tag{36}$$

Then taking into account (9) we have after a GWT

$$\tilde{S}_{|\nu} = e^{n\lambda} S_{|\nu}.  \tag{37}$$

Thus, (36) is a gauge-covariant operation. Further, for a vector $V^{\nu}$ with the Weylian power $\Pi(V^{\mu}) = n$ we have the Weylian derivative with $\Gamma^{\lambda}_{\mu\nu}$ given by (3)

$$\nabla_{\nu} V^{\mu} = V^{\mu}_{,\nu} + V^{\sigma} \Gamma^{\mu}_{\sigma\nu}.  \tag{38}$$

This is coordinate-covariant but it is not covariant with respect to WGT, unless $n = 0$. Instead of (38) we can define a covariant-covariant (covariant with respect to both CT and WGT) derivative

$$\overset{*}{\nabla}_{\nu} V^{\mu} \equiv \nabla_{\nu} V^{\mu} - nV^{\mu} w_{\nu} = V^{\mu}_{,\nu} + V^{\sigma} \Gamma^{\mu}_{\sigma\nu} - nV^{\mu} w_{\nu}.  \tag{39}$$

One easily proves that

$$\overset{*}{\nabla}_{\nu} \tilde{V}^{\mu} = e^{n\lambda} \overset{*}{\nabla}_{\nu} V^{\mu}.  \tag{40}$$

Introducing a new connection

$$\overset{*}{\Gamma}^{\lambda}_{\mu\nu} \equiv \Gamma^{\lambda}_{\mu\nu} - n\delta^{\lambda}_{\mu} w_{\nu},  \tag{41}$$

one can rewrite (39) as

$$\overset{*}{\nabla}_{\nu} V^{\mu} = V^{\mu}_{,\nu} + V^{\sigma} \overset{*}{\Gamma}^{\mu}_{\sigma\nu}.  \tag{42}$$

Equation (42) justifies introducing the following law of parallel displacement (cf. [12])

$$\overset{*}{d} V^{\mu} = -V^{\sigma} \overset{*}{\Gamma}^{\mu}_{\sigma\nu} dx^{\nu}.  \tag{43}$$

With (9) one obtains $\overset{*}{d}\tilde{V}^{\mu} = e^{n\lambda}\overset{*}{d}V^{\mu}$, so that (43) is a *gauge-covariant displacement*.

The change in the length of a vector after the gauge-covariant displacement (43) can be calculated from $\overset{*}{d}(V)^2 \equiv \overset{*}{d}(g_{\mu\nu}V^{\mu}V^{\nu}) = 2g_{\mu\nu}V^{\mu}\overset{*}{d}V^{\nu} + V^{\mu}V^{\nu}g_{\mu\nu,\sigma}dx^{\sigma}$. Making use of (41), (43), (3) one obtains

$$\overset{*}{d}(V)^2 = 2V^2(n+1)w_{\sigma}dx^{\sigma}, \qquad (44)$$

$$\overset{*}{d}V = V(n+1)w_{\sigma}dx^{\sigma}. \qquad (45)$$

This formula plays an important role in the W-D framework, as it is derived from the derivative (39); the latter being covariant with respect to both, CT and WGT.

According to (45) the change in length depends on the Weylian power of the vector. For example, if $n=0$ we get from (45) $\overset{*}{d}V = Vw_{\sigma}dx^{\sigma}$, which agrees with equation (2). This is not surprising, as in section 2. we assumed $\Pi(V^{\mu})=0$. However, an interesting result follows from (45) when

$$\Pi(V^{\mu}) = -1; \quad n = -1. \qquad (46)$$

In this case we obtain

$$\overset{*}{d}V = 0. \qquad (47)$$

Following Rosen [12] we will call a vector with power $\Pi(V^{\mu})=-1$ a **standard vector**. From (46) and $\Pi(g_{\mu\nu})=2$, one obtains immediately $\Pi(V_{\mu})=1$ and $\Pi(V)=0$. Thus, the length of a standard vector is a gauge-invariant quantity, and in the process of parallel displacement it does not change.

If atoms are characterized by standard vectors, their intrinsic properties will remain unchanged in the process of parallel displacement, and such particles may serve as measuring standards corresponding to the atomic gauge of Dirac [6].

## 6. ACTION INTEGRAL AND FIELD EQUATIONS

Above some possibilities of describing the interior geometry of a measuring-atom were discussed. In the present section we consider in brief some details concerned the action and field equations.

It must be emphasized that for any model, the physical reality outside the atom is that of the classical Weyl-Dirac theory with the Dirac action (cf. [6], [3])

$$I = \int \left[ W^{\lambda\mu}W_{\lambda\mu} - \beta^2 R + 6\beta_{,\lambda}\beta_{,\underline{\lambda}} + 2\Lambda\beta^4 + L_{\text{matter}} \right](-g)^{1/2} d^4x \quad . \tag{48}$$

In (48) $R$ is the Riemannian curvature scalar, $\Lambda$ stands for the cosmological constant and an underlined index is raised with $g^{\mu\nu}$. From (48) one derives (cf. [3], [12]) the field equations and the equation of motion of a particle, having rest mass $m$ and electric charge $e$:

$$G^{\mu\nu} = -\frac{8\pi}{\beta^2}\left(M^{\mu\nu} + T^{\mu\nu}\right) + \frac{2}{\beta}\left(g^{\mu\nu}\beta_{;\alpha;\underline{\alpha}} - \beta_{;\underline{\mu};\underline{\nu}}\right)$$
$$+ \frac{1}{\beta^2}\left(4\beta_{,\underline{\mu}}\beta_{,\underline{\nu}} - g^{\mu\nu}\beta_{,\alpha}\beta_{,\underline{\alpha}}\right) - g^{\mu\nu}\Lambda\beta^2 \quad , \tag{49}$$

$$W^{\mu\nu}_{\;;\nu} = 4\pi J^{\mu} \quad , \tag{50}$$

and

$$\frac{dU^{\mu}}{ds} + \left\{{}^{\mu}_{\lambda\nu}\right\}U^{\lambda}U^{\nu} + (U^{\mu}U^{\nu} - g^{\mu\nu})\frac{\beta_{,\nu}}{\beta} = \frac{e}{m}U_{\nu}W^{\mu\nu} \quad . \tag{51}$$

In equations (49) and (50), $M^{\mu\nu} \equiv \frac{1}{4\pi}\left(\frac{1}{4}g^{\mu\nu}W^{\lambda\sigma}W_{\lambda\sigma} - W^{\mu\lambda}W^{\nu}_{\;\lambda}\right)$ is the energy-momentum density tensor of the electromagnetic field, $T^{\mu\nu}$ is that of matter, and $J^{\mu}$ is the electric current density vector.

At the same time, within the measuring-atom the physical reality is described differently.

1. If one adopts the bubble model of Papini and Wood (cf. section 3), one has inside the atom $w_\mu = 0$, so that $W_{\mu\nu} = 0$, and hence there is a Riemannian space with Einstein's GTR holding.

2. One can choose the model presented in section 4. Here, inside the measuring-atom one has a space characterized by $g_{\mu\nu}$ and the connection $\underset{3}{\Gamma}^\lambda_{\mu\nu}$ (cf. (29)) without nonmetricity (consequently without the gauge function) but with torsion. In this case the Maxwell field tensor is the divergence of torsion (cf. (34))

$$W_{\mu\nu} \equiv w_{\mu;\nu} - w_{\nu;\mu} = 2\underset{3}{\Gamma}^\lambda_{[\nu\mu];\lambda} . \tag{52}$$

Let us write the action integral as [1]

$$\underset{3}{I} = \int \left( W^{\lambda\mu} W_{\lambda\mu} - K\left(\underset{3}{\Gamma}^\lambda_{\mu\nu}\right) + 2\Lambda + L_{\text{matter}} \right) \sqrt{-g}\, d^4x , \tag{53}$$

with $K\left(\underset{3}{\Gamma}^\lambda_{\mu\nu}\right)$ standing for the curvature scalar formed from the connection $\underset{3}{\Gamma}^\lambda_{\mu\nu}$. By a straightforward calculation one obtains

$$K\left(\underset{3}{\Gamma}^\lambda_{\mu\nu}\right) = R\left(\{^\lambda_{\mu\nu}\}\right) - 6w^\sigma_{;\sigma} + 6w^\sigma w_\sigma . \tag{54}$$

Substituting this into (53) and discarding the divergence term one obtains the action

$$\underset{3}{I} = \int \left( W_{\lambda\sigma} W^{\lambda\sigma} - R - 6w^\lambda w_\lambda + 2\Lambda + L_{\text{matter}} \right) \sqrt{-g}\, d^4x , \tag{55}$$

with $W_{\mu\nu}$ given by (52). The field equations derived from (55) take the form

$$G^{\mu\nu} = -8\pi\left(M^{\mu\nu} + T^{\mu\nu}\right) + 3g^{\mu\nu} w^\lambda w_\lambda - 6w^\mu w^\nu - g^{\mu\nu}\Lambda , \tag{56}$$

$$W^{\mu\nu}_{;\nu} = 4\pi J^\mu - 3w^\mu . \tag{57}$$

---

[1] A different, interesting approach - the $\lambda$-symmetry of the action integral - is developed and discussed in the works of A. Einstein and B. Kaufmann [16] as well in a paper by E. I. Guendelman [17].

Now, from (31) one has $w_\nu = \frac{2}{3} \Gamma^\lambda_{3\,[\lambda\nu]}$, so that equation (57) contains a Proca [18] term involving torsion.

For a moment let us adopt the naïve model of Bohr's atom. Then, in regions free of electric current we have

$$W^{\mu\nu}_{\;;\nu} + 3w^\mu = 0. \qquad (58)$$

This is the covariant form of the Proca equation for a vector boson field. From (58) follows the condition

$$w^\mu_{\;;\mu} = 0, \qquad (59)$$

which may be considered as a reminder of the Lorentzian gauge condition. From (58), (59) one obtains

$$w^\mu_{\;;\nu;\underline{\nu}} + w^\nu R^\mu_\nu + 3w^\mu = 0 . \qquad (60)$$

and if the curvature is negligible one can write

$$w^\mu_{\;;\nu;\underline{\nu}} + 3w^\mu = 0. \qquad (61)$$

This describes a vector field, which from the quantum mechanical standpoint is represented by bosons of spin 1 and of mass $m = \frac{\sqrt{3}\,\hbar}{c}$ ($m \approx 2.24 \times 10^{-21} m_e$).

Now, let us go back to equations (55) – (57). Taking into account the relation $w_\nu = \frac{2}{3} \Gamma^\lambda_{3\,[\lambda\nu]}$ and (52), we conclude that torsion is involved in the action integral as well in the in the field equations.

3. One can assume that the measuring-atom is characterized by Rosen's standard vectors (cf. section 5). By this choice both regions, the interior and the exterior are described by equations (48) – (51).

# 7. CONCLUSION

Recently [1] a generalization of Wesson's Induced Matter theory [2] was presented. The 5-dimensional bulk was provided with a Weyl geometric framework, and the 5-D field equations were derived from an action principle. As a result on the 4-D brane a Weyl-Dirac framework was obtained, with field equations containing terms induced by the bulk.

If one is looking for a geometric description of classical gravitation and electromagnetism the Weyl geometry is attractive because it provides a geometric vector $w_\mu$ that can be interpreted as the electromagnetic potential vector. The problem that faces us in Weyl's geometry is the nonintegrability of length. We can overcome this obstacle, if following Dirac [6], we adopt the standpoint that the geometry in the interior of atoms, which serve as measuring standards, differs from the geometry describing the exterior. In the present paper various models describing the interior of atoms are presented. With the outcome of the present note the Weyl-Dirac theory may be considered as a suitable framework for describing physical phenomena.